# Manuscript Title:

# Ultrahigh-sensitivity high-linearity photodetection system using a low-gain avalanche photodiode with an ultralow-noise readout circuit

**Authors:** Makoto Akiba, Mikio Fujiwara, and Masahide Sasaki

**Address:** National Institute of information and Communications Technology, Basic and Advanced Research Department, 4-2-1 Nukui-Kitamachi, Koganei,Tokyo 184-8795, Japan

**Abstract**

A highly sensitive photodetection system with a detection limit of 1 photon/s was developed. This system uses a commercially available 200-µm-diameter silicon avalanche photodiode (APD) and an in-house-developed ultralow-noise readout circuit, which are both cooled to 77 K. When the APD operates at a low gain of about 10, it has a high-linearity response to the number of incident photons and a low excess noise factor. The APD also has high quantum efficiency and a dark current of less than 1 e/s at 77 K. This photodetection system will shorten the measurement time and enable higher spatial and wavelength resolution for near-field scanning optical microscopes.

OCIA codes: 040.3060, 040.3780, 040.5160, 040.5350, 040.6040, 040.6070

Studies on cooled APDs have shown that their sensitivity operating at low temperatures exceeds that of photomultiplier tubes (PMTs) at near-infrared wavelengths because of their high quantum efficiency.[1,2] When an APD is cooled to low temperatures, its dark current and current noise decrease significantly,[1-3] while the quantum efficiency decreases only slightly.[4] As the APD noise decreases at low temperatures, amplifier noise becomes the dominant factor that limits the performance of the APD.

Generally, in a pulse height distribution of the APD output signals for single photons, the probability that lower-height pulses are generated is greater than that for higher-height pulses.[2,3,5] In photon counting mode, lowering the discriminator threshold thus results in a higher quantum efficiency. Amplifier noise, however, limits the threshold level. In linear mode, the fluctuation of the avalanche gain prevents lower-level light detection. Since the fluctuation also decreases as the gain is decreased, the sensitivity can be further improved by using a low-noise amplifier with an APD operating at a low gain.[6,7]

In light of the above issues, we developed a photodetection system with a low-gain APD operated in linear mode to detect ultralow-level light signals (in the order of 10 photons/s) with high sensitivity and high linearity. We aim to apply it to measurements with near-field scanning optical microscope (NSOM), where both a short-pulse and a continuous-wave signals must be detected. High-gain APDs in photon counting mode cannot be used for the intensity measurements of a short-pulse signal because they cannot discriminate the number of simultaneously incident photons. When APDs are operated in linear mode, they can be used for the measurements; however, in the case that a conventional amplifier is used, the photodetection limit of APDs in linear mode is more than double that of APDs in photon counting mode because of large excess noise.[1] A visible light photon counter (VLPC) is an alternative detector for the intensity measurements of a short-pulse signal.[8-10] However, a VLPC suffers from a high dark-count rate when it is used for detecting longer pulses or continuous-wave signals.

The readout circuit we used for this work was a kind of charge-sensitive amplifier, a so-called capacitive transimpedance amplifier, and was the same as that used by Akiba and Fujiwara.[11] This readout circuit has ultralow current-noise at low frequencies and a low input capacitance of 1 pF. The APD used was a 200-μm-diameter bare chip, model M6612, made by Matsusada Precision Inc. The capacitance of the APD was about 1 pF at DC bias voltages of about 90 V. The readout circuit works at an ultralow noise level at this capacitance,[11] so the APD can be operated at very low gains of about 10. The readout circuit and the APD were mounted on a $SiO_2$ glass platform on the work surface of a liquid nitrogen cryostat. A 635-nm light-emitting diode (LED) was used as the light-pulse source for measurements of the performance of our photodetection system. The LED was mounted in a box with a pinhole to attenuate the intensity of the light from it. The box was directly attached to the cryostat outside wall, covering the whole cryostat window, so that the light in the room was kept out of the cryostat. The schematic response of the readout circuit to the light pulses is shown in Fig. 1. The readout noise of the circuit, which was measured under the condition given in Fig. 1 without light pulses, was seven electrons. The number of photoelectrons created in the APD was calibrated by operating the APD as an ordinary photodiode at zero bias voltage, and integrating the photoelectrons created by 2,000 light pulses. Figure 2 shows the mean gain measured for the light pulses that, on average, create a single photoelectron.

The electron-number distributions at the APD anode for three mean numbers of photoelectrons at gains of 10.8 and 31.1, which correspond to bias voltages of 92.2 and 99.5 V, respectively, are shown in Fig. 3(a) and (b). For comparison, the probability distribution of the electron number, $N(x)$, was estimated from the following expression,

$$N(x) = \frac{1}{\sqrt{2\pi}\sigma} \sum_l P_n(l) \exp(-(x - Ml)^2 / 2\sigma^2), \tag{1}$$

where $l$ is the photoelectron number, $P_n(l)$ is the Poisson distribution of the average of $n$, $\sigma$ is the standard deviation of readout noise, where $\sigma = 7$, and $M$ is the mean gain at the bias voltage. In this expression, it is assumed that the output signal of the APD is proportional to the mean gain and the photoelectron number. The Gaussian function represents peak broadening by the readout noise, but this expression does not include the noises associated with the APD. The estimated distributions are represented as the dashed lines in the figures. At the gain of 10.8, there is little difference between the measured and estimated distributions. This result indicates that the noise associated with the APD at that gain was smaller than the readout noise, and that the APD has a high linearity with respect to the number of incident photons. This feature was seen in all the cases tested for the eight values of the mean gain up to 10.8. On the other hand, for the two points above the gain of 17, the noise associated with the APD was found to be dominant, since the electron-number distributions are far from the estimated ones as seen in Fig. 3(b).

The excess noise factors of the APD and the total excess noise factors that include both readout noise and APD noise were evaluated from the electron-number distributions for six mean numbers of photoelectrons (Fig. 4). At a gain of 31.1, the two excess noise factors are almost the same, because the noise of the APD is dominant. The excess noise factors at a gain of 10.8 are much smaller than those predicted by McIntyre's theory,[5] which assumes a continuous multiplication process. In fact, the theory predicts a lower limit of 1.91, according to the following formula for electron photocurrent injection:

$$F = k\,M + (2 - 1/M)(1 - k), \qquad (2)$$

where $k$ is the hole-to-electron ionization rate ratio, which is very small, i.e. about 0.006 for a silicon APD.[3] The over-estimation of the excess noise factor has been pointed out by several authors. Hayat et al. introduced the dead-space model to show that the reduction of the excess noise factor is expected for an APD with thinner active layers.[12] McIntyre and others developed the history-dependent theory and successfully explained the experimental excess noise factors.[13,14] Even in light of these works, our values seem to be very small for a gain of about 10. For further analysis of the excess noise factor, we must know the structure of the APD in detail, which is unfortunately unavailable at present, so we must leave that analysis for future study.

The dark current of the APD can be determined from the time variation of the output voltage of the readout circuit when the light pulse is not input. The output-voltage drift caused by the voltage or current drifts at various parts in the circuit, however, disturbs the measurement of the dark current. For example, the drifts of the input offset voltage of the operational amplifier or the drifts of the drain current of readout junction field-effect transistor may be responsible for the output-voltage drift. If the time variation of the output voltage results from the dark current of the APD, the time variation depends on the bias voltage or the gain of the APD.[3] In our photodetection system, the time variation was independent of the bias voltage. As a result, an upper-limit dark current of 1 e/s was obtained from the maximum variation of the output voltage in 1 s.

The quantum efficiency of the APD was indirectly determined from the room-temperature quantum efficiency given in the manufacturer's catalog. The APD was operated as an ordinary photodiode at zero bias voltage, and the output current at 77 K was compared with the output current at room temperature. The quantum efficiency dropped by 8% at 77 K. Since the room-temperature quantum efficiency at a wavelength of 635 nm is at least 69%, the 77-K quantum efficiency of our APD was estimated to be greater than 61%. From the quantum efficiency, a readout noise of 7 e, and a gain of 10.8, the photodetection limit of our system for a light-pulse interval of 50 ms is predicted to be 1.1 photons. For a continuum light source, the limit increases up to 1.5 photons in a correlated double sample with a 50-ms sampling interval. Furthermore, dark noises corresponding to light-pulse intervals of 1 s, 1 ms, and 0.01 ms were measured. From these noises, the photodetection limits are estimated to be 0.8, 1.3, and 10 photons. From these values of photodetection limits, the time resolution of our system under a photodetection limit of one photon is found to be low. However, since the readout circuit is charge-sensitive, the integrated charge can be readout after a light pulse has been incident on the APD (see Fig. 1). When the interval of the light pulse is larger than 1 ms, our system can detect a short light-pulse at a photodetection limit of one photon.

In conclusion, our photodetection system allows more precise measurements of the intensity of ultralow-level light signals than conventional photon counters because of its higher quantum efficiency and lower dark current. This system has high sensitivity and high linearity so it can be applied to a wide range of low-light-level experiments, such as those using a short-pulse laser or a continuous-wave laser, and experiments with no background light or high background light.

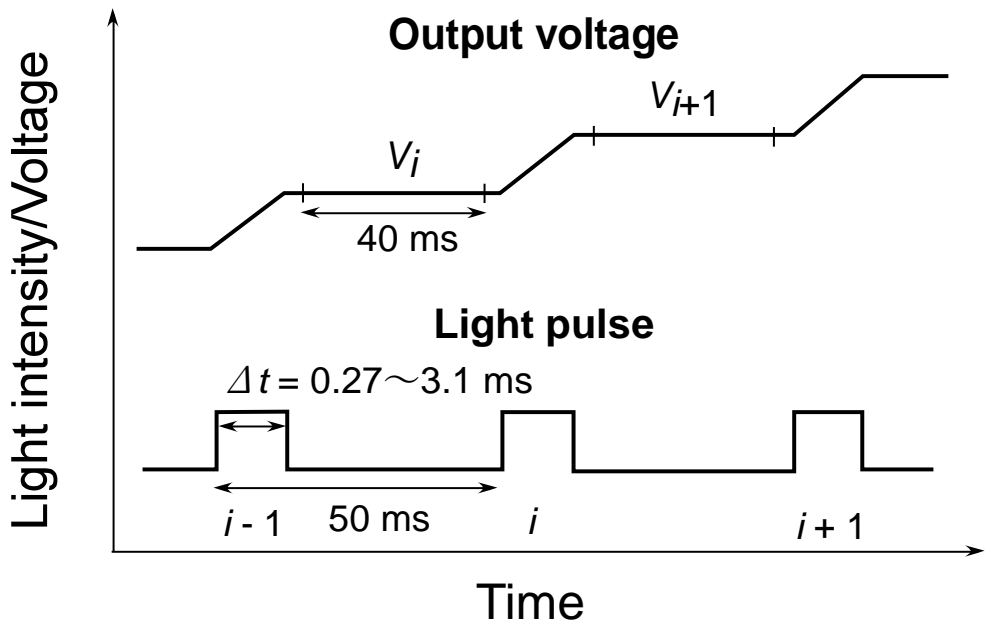

Figure 1. Schematic response of the readout circuit to the light pulses. The interval of the light pulses is 50 ms. The average intensity of the LED light was controlled by varying the pulse width from 0.27 ms for one photoelectron to 3.1 ms for 10 photoelectrons. $V_i$ is the $i$th voltage averaged for a period of 40 ms in the interval of light pulses. The integrated charge $Q_i$ of $i$th output signal for the $i$th light pulse is given by $Q_i = C_f(V_{i+1} - V_i)$, where $C$ is a feedback capacitance.

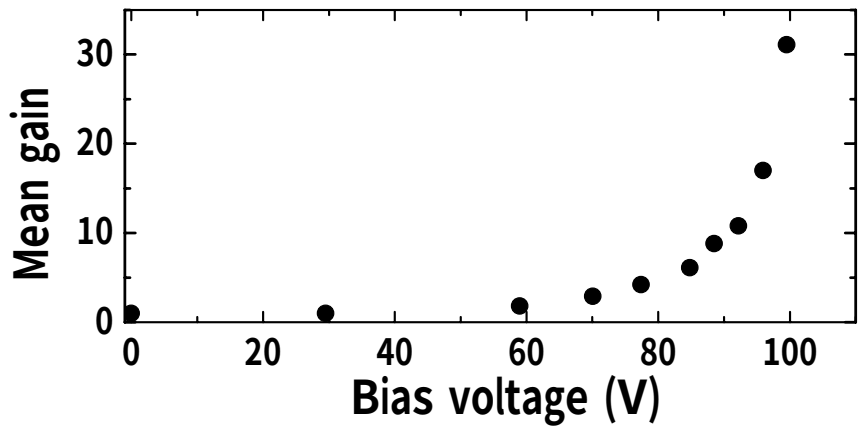

Figure 2. Mean gain of APD as a function of reverse bias voltage

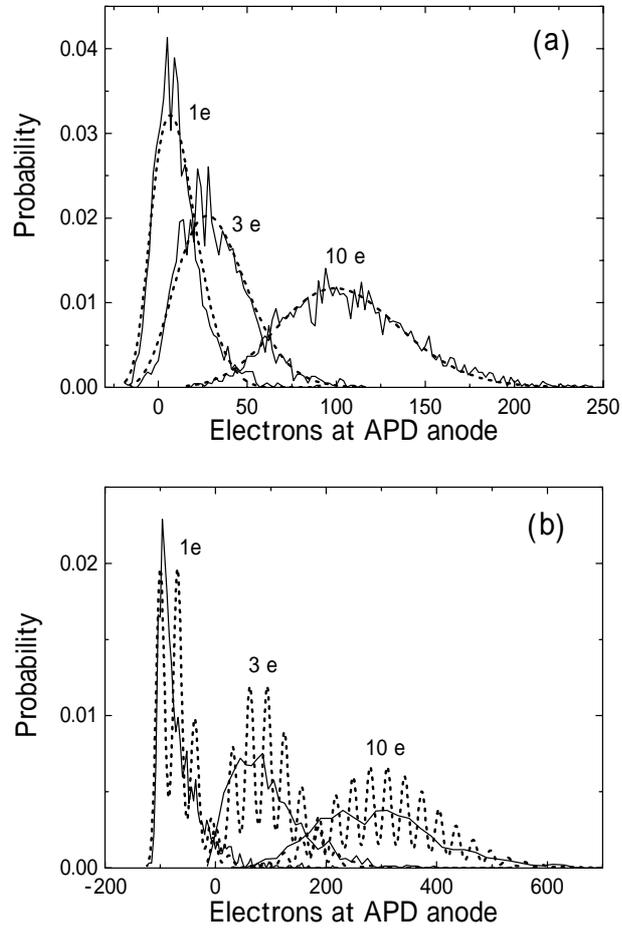

Figure 3a-b Measured electron-number distributions at the APD anode (solid lines) for mean numbers of photoelectrons equal to 1, 3, and 10 (a) at a gain of 10.8 and (b) at a gain of 31.1. Dashed lines are calculated distributions. The distributions for a mean photoelectron number of 1 in (b) are displaced horizontally by –100 for clarity.

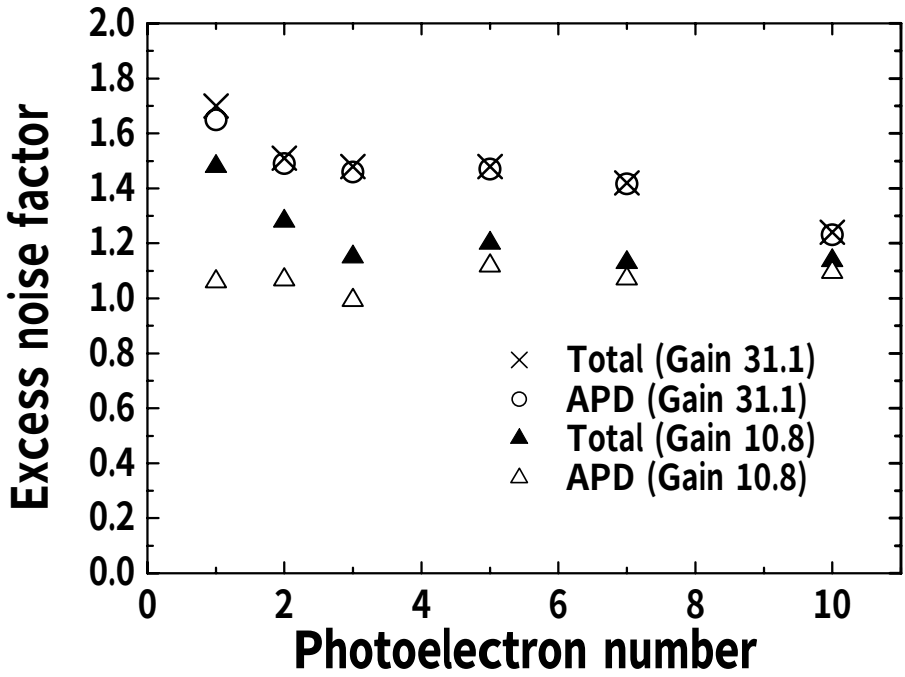

Figure 4. Dependence of the excess noise factor on the number of photoelectrons at gains of 10.8 and 31.1. The total excess noise factor includes both readout noise and APD noise.